\documentclass[twocolumn,floatfix,aps]{revtex4}

\begin{document}
\title{Reply to the Comment [arXiv:0810.3247v1] by G. L. Klimchitskaya et al. on "Application of the Lifshitz theory to poor conductors"  }

\author{Vitaly B. Svetovoy}
\affiliation{MESA+ Research Institute, University of Twente, PO 217,
7500 AE Enschede, The Netherlands}

\date{\today}

\begin{abstract}
It is shown that the claims expressed in the Comment
arXiv:0810.3247v1 against my paper Phys. Rev. Lett. {\bf 101},
163603 (2008) are obviously wrong or not essential.
\end{abstract}

\maketitle

The authors of the Comment \cite{Kli08} on my Letter \cite{Sve08c}
put forward a number of points, which they consider as erroneous. To
my opinion the Comment does not appear to be scientifically valid.
Below it is shown that the claims are obviously wrong or not
essential to my Letter.

Before going into details let me enumerate the blames expressed in
the Comment to summarize the statements and for the convenience of
the following references.

\begin{enumerate}

    \item (i) Spatial dispersion is taken into account approximately;
    (ii) one can define the nonlocal dielectric function only for infinite medium;
    (iii) specular reflection is not a good approximation.
    \item Uncertainty in $n$, $\Delta n =0.4\times 10^{19}\
    cm^{-3}$, was determined in \cite{Che07} at 95\% confidence level, but the
    author of the Letter did not indicated it.
    \item Nonlocal approach does not agree with the force measured
    in \cite{Dec07}.
    \item Temperature dependence of the charge concentration for
    ionic conductors is incorrect. As the result the Nernst
    theorem is broken for ionic conductors.

\end{enumerate}
The claim 4. is equally applied to Refs. \cite{Pit08,Dal08}.

{\bf 1}. (i) Any physical theory can describe the nature only
approximately. Indeed, the random phase approximation gives only
approximate expression for the dielectric function, which works well
at the wave numbers $k\ll k_F$, where $k_F$ is the Fermi wave
number. In the Casimir problem typical wave numbers are $k\sim
1/a\ll k_F$, where $a$ is the distance between bodies. All that was
explained in the Letter and the authors of the Comment did not
propose anything to overturn my argumentation.

(ii) The nonlocal dielectric function can be easily defined for
infinite medium but in reality all bodies have boundaries. This is
true in general for all problems where spatial dispersion is
important. The nonlocal response for infinite medium can be used to
build a correct solution for semi-infinite body using methods
developed in the theory of anomalous skin effect.

(iii) In the Letter I have chosen the specular reflection of
electrons on the body surfaces as the most simple boundary
condition. Diffusive and partly specular conditions were discussed
in the literature. The preferable condition depend on the quality of
the surface. If de Broglie wavelength of electrons is large in
comparison with the root-mean-square roughness of the surface then
the electron reflection can be considered as specular. This
condition is true for semiconductors used in microtechnology and for
metallic films with low roughness ($\lesssim 1\ nm$). The fact that
the Debye screening was successfully reproduced in the Letter also
shows that this boundary condition is reliable.  It was demonstrated
many times in the literature that for partly specular or diffusive
boundary conditions the qualitative conclusions will be the same.
For example, the impedance of anomalous skin effect varies only 10\%
when the boundary condition continuously changes from specular to
diffusive \cite{Har66}.

In the Comment we find: "for spatially dispersive materials the
scattering of charge carriers is neither specular nor diffuse
\cite{Fol75}." The authors of the Comment give wrong interpretation
of the paper \cite{Fol75}, where scattering of electrons was not
even discussed. The paper was devoted to a macroscopic approach to
spatial dispersion that is justified at some specific conditions.
This approach is less general than the microscopic approach
(scattering of electrons).

{\bf 2}. In the Letter I did not indicate that in the expression
\cite{Che07} $n = (2.1\pm 0.4)\times 10^{19}\ cm^{-3}$, the error
was determined at 95\% confidence level (CL). I could not even do
this because it was not mentioned anywhere in Ref. \cite{Che07}.
Equation (16) in this paper was used to estimate the error. The
input values of $\tau$ and $w$ for this equation were defined with
one standard deviation error, $1\sigma $ (68\% CL). In equation
(17), above it, below it, or anywhere else it was not indicated that
the error is not standard. The paper \cite{Che07} is a document
which I cannot change.

Anyway it is clear from Fig. 1(a) of the Comment that deviation of
the nonlocal theory from the experiment is hardly convincing. Even
smaller deviations realized for the exciting power $P=8.5\ mW$ as
one can estimate from Fig. 1(b) in the Letter reducing the width
of the gray stripe. Deviation of a physical theory from
experiments on the level of $1\sigma $ cannot be considered
seriously.

{\bf 3.} For good metals at room temperature the nonlocal effects
are negligible. It is clear from Eq. (13) and Ref. [16] in the
Letter. This is because the Thomas-Fermi screening length for good
metals is very short, $k_D^{-1}\sim 1\AA$. That is why the gray
stripe shown in Fig. 1(b) of the Comment has nothing to do with the
spatial dispersion effect. Moreover, the reference to the experiment
\cite{Dec07} is irrelevant at all since in this experiment the
optical properties of used Au films were not measured. The force was
calculated for some imaginary material (perfect single crystal in
far infrared (IR) and some film in IR and higher frequencies) and
then compared with the experiment. Recently it was shown
experimentally that this procedure is wrong \cite{Sve08b}: optical
properties of different Au samples are significantly different.

{\bf 4.} The authors of the Comment state that density of charges
for ionic conductors does not depend on the temperature. They refer
to the paper \cite{Tom98} where the opposite is stated (see the last
paragraph in Sec. 5): "... the conductivity change with temperature
is accompanied by the change in the charge carrier concentration."
It is clear from general physical consideration that at low
temperature the density of charges must disappear exponentially. The
charges in an ionic conductor appear due to dissociation of neutral
molecules. At finite $T$ at equilibrium the charge density $\sim
e^{-E_a/T}$. When $T$ decreases these charges are neutralized by the
same exponential law if one follows the equilibrium state. In
absence of equilibrium the entropy can stay finite at $T=0$.

\end{document}